\documentclass[12pt,a4paper,final]{iopart}
\usepackage{iopams}
\usepackage{graphicx}
\usepackage[breaklinks=true,colorlinks=true,linkcolor=blue,urlcolor=blue,citecolor=blue]{hyperref}

\usepackage{epsfig}
\usepackage{amssymb}
\usepackage{dcolumn}
\usepackage{bm}

\def\be{\begin{eqnarray}}
\def\ee{\end{eqnarray}}
\def\ba{\begin{array}}
\def\ea{\end{array}}

\begin{document}
\title[Gestalt versus Heisenberg]{Gestalt Principles re-investigated within Heisenberg uncertainty relation}
\author{S. Iyikalender and A. Siddiki}
\address{ Mimar Sinan Fine Arts University, 34380 Sisli, Istanbul, Turkey}

\ead{afifsiddiki@gmail.com}
\vspace{10pt}
\begin{indented}
\item[]March 2017
\end{indented}
\begin{abstract}
Perception, sensation and re-action are central questions both in Psychology, Arts, Neurology and Physics. Some hundred years ago, believed to start with Wertheimer, researchers and artists tried to classify our human being "understanding" of Nature, in terms of \emph{Gestalt} principles. During same period \emph{Quantum} mechanics were developed by Schroedinger, Heisenberg, Dirac, Majorana and others. In this work we briefly summarize the basic concepts of these two approaches and try to combine them at a simplistic level. We show that, perception and sensation can be handled within electrical signal processing utilizing Fourier transformation, which finds its counter-part in quantum mechanics.  
\end{abstract}

\section{Introduction}
During the evolution, we as \emph{Homo Sapines} have reached, as far as we know, the peak point of understanding the nature \cite{Moses:5000,Jeusus:0,muhammed:652}. Although, loosing our respect to it and raping this beauty. Our nerve system together with our brain, we could process and proceed things happening around \cite{Beethoven:7, Bruno:G, Goedel:t}. However, it is still a major question "how" we conceive, convolve and react the information coming from \emph{outside}. For seeing, it is clear that an electro-magnetic signal stimulates our eye cells. This input is converted to an electrical signal via Fourier transform and is conveyed by neurological wires to the acceptor center. By "center" we mean the \emph{orthodox} understanding, which as authors barely understand. 

Exactly 105 years ago, Wertheimer \cite{Wertheimer:12} summarized the basics of our perception called the \emph{Gestalt Principles}. He states that, there should be certain rules in our perception and sensation. However, in the following years discussion was pushed forward by many many researchers. These principles are used by  psychologists, artists and others. In spite of all the efforts, the main mechanism lying behind is still under debate. Recently, in a mind blowing pioneering work \cite{WAGEMANS:12} (and the references given there) mathematical and neurological background of Gestalt principles are discussed in detail, deep and firmly. We think that there is sufficient space to improve on the re-vised \emph{Gestalt Principles}, utilizing neurological findings and quantum mechanical approach.

In this letter, we provide an approach that may interconnect the relation between gestalt principles (GPs) and uncertainty relation (UR) via Fourier transformation (FT). First, we very briefly remind the readers GPs, then re-introduce discrete FT and finally review UR. Second we, show some of our simple but fundamental results and relate GPs to UR.         
\section{Gestalt Principles}
Approximately 2.5 Milion years ago our ancestors woke up, i.e. stand on their legs. They were called \emph{Homo Erectus}, the erected ape. Their vision were much greater than the monkeys, then. Nobody knows, why they erected? It is out of interest for us ``why?", our target is to answer ``how?". We would like to emphasize that, the authors are not sure that it was a good idea to erect, if one observes the contemporary situation of human beings in modern times. 

This erection lead more information due to the new point of view. How do we and other eyed animals percept information by grouping and finally analyze is not clear, yet. As a step forward, process and react this premature information is even more complicated. 

GPs of grouping (perception in a limited case), provided somehow consistent, coherent and may be logical ground for our visualization. Those are \cite{GSp:2017}:  
\begin{itemize}
	\item Proximity
	\item Similarity
	 \item Closure
	 \item "Good" Continuation
	 \item Common "Fate"
	 \item Good "Form"
\end{itemize}
  
The first three principles are rather clear and scientifically (in mathematics, physics and neurology Etc.) digestible. However, the last three are questionable.

Just as an (simple) example we show in Fig.1 the proximity effect, together with non-grouping. GPs imposes that due to proximity effect, perception is \emph{better}.  We will show that, this statement is not true, at least signal wise. 

\section{Fourier Transformation}
 In applied mathematics, counting is fundamental. Historically, as far as we know, it all started making scratches on wood mapping \emph{number} of sheep leaving or coming back to yard. Fig 2. Hence, there were two different worlds. The reality was singular, sheep to be fed and to feed yourself. To be simple as possible, mathematicians map the real world to some symbols like scratches or numbers Etc. Moreover, they find some smart way to relate two different realities with each other: observables and observers. 
 
 Now, many many years later Fourier realized that, if one takes any function (a special case of relation) it is possible to express this function in terms of already known functions. One of the well known are Cosine and Sine, periodically repeating functions. Therefore, one can expand in principle a given electro-magnetic signal (function) in bases of these. Fig 3. \cite{FT:14}. Mathematically;
 \begin{equation}
f(k)=\int_{-\infty}^{\infty}f(x)e^{-2\pi i x k}dx,
 \end{equation}
 here $x$ is a real number and $k$ is the dual space. Similar to mapping between scratch and sheep. For the interested reader we suggest to look at the duality concept from Mevlana, Goethe, Hegel and Marx \cite{duality}.
 
 In any case, we hope it is now clear to an untrained reader that there are two different, however, equivalent sets/spaces, which can be related to each other by Fourier transformation. 
 
Unfortunately and mathematically ugly, there is a \emph{funciton} called delta, $\delta(x,k)$, i.e. relates $x$ and $k$ spaces. Presented as;

\begin{equation}
\delta (x-a)=\frac{1}{2 \pi}\int_{-\infty}^{\infty}dk\quad \cos (kx-ka).
\end{equation}
Here $a$ is the period, i.e. repeating number. By period $a$ it is meant that, the repeating distance between peak points of the function, as shown in Fig.3. The mathematical ugliness of $\delta$ function comes from the fact that, in one space ($x$ or $k$) it diverges to infinity, meanwhile in the dual space tends to zero. More disgustingly, the area below this function (integral) is constant always and equates to 1.

In the next Section, basics of Uncertainty relation will be re-introduced, while the FT in perception is strongly related with it.

\section{The Very Basics of Quantum Mechanics\label{sec:toy_model}}

Early in 1899, Max-Planck realized that the very well known \textit{classical} mechanics and statistics does not explain the observed facts, which are obviously repeatable experimentally. Hence he, as a genius, developed a novel mathematical formulation to explain the observed in nature, black body radiation. However, his work in the beginning, was not more than collecting ideas which already existed. Un (or) fortunately, there was the ``catastrophe" at Ultra-Violet radiation \cite{Loudon:00}, that can not be explained by classical means. He had to invent a new formulation,        
\begin{equation}
\frac{<N_i >}{N}=\frac{g_i}{e^{(\epsilon_{i_{(n)}} -\mu)/kT}},
\end{equation}
which essentially expresses that light is composed of quantized particles, as Newton stated couple of years ago \cite{Newton}. In this formula, $i$ counts the state of particle number, $\mu$ is the chemical potential determining the probability to include another particle to the system, $T$ is the so called temperature, and $k$ represents Boltzmann constant. We, added $(n)$ subindex to emphasize the discreetness of energy level, similar to $g_i$, representing the Spin degree of freedom. Here, $N$ is the total number of particles and, $< N_i >$ is the expected number of particles in $i\quad th$ state.  

The report by Max-Planck paved the way to quantum mechanics \cite{Planck}. However, it was Heisenberg who formulated QM independent of physical nature just grounding on triangle inequality;
\begin{equation}
|<\mathbf{u,v}>|^2 \leq <\mathbf{u},\mathbf{u}>. <\mathbf{v},\mathbf{v}>
\end{equation}
where $\mathbf{u}$ and $\mathbf{v}$ are the two corners of a triangle, as shown in Fig 4. Here $<.,.>$ is the inner product.

At the last step, we re-present the Heisenberg uncertainty relation;
\begin{equation}
[\hat{x},\hat{p}]{\psi(x)}\geq i\hbar \quad \psi(x).
\end{equation}
In fact, wave function $\psi(x)$ also depends on \emph{time}, however, it is pedagogically ignored usually. To be brief, it is the fact that mathematically, using Schwartz inequality one can prove that;
\begin{equation}
[\hat{A},\hat{B}]=\hat{A}\hat{B}-\hat{B}\hat{A},
\end{equation}
$\hat{A},\hat{B}$, are operators that can be replaced by coordinate and momentum operators $\hat{x}$ and $\hat{p}$  which does not commute with each other. For a uninterested and ordinary reader, the two perpendicular sides of a triangle can not be larger than the third line that connects the open edges of the sides.
Finally, while knowing it is a long path to investigate \emph{perception, sensation and re-action} we have found an interconnection between GPs, FT and UR. This it requires much more effort, than presented here.  

Now, we able to discuss Gestalt principles within Heisenberg \cite{Heisenberg} uncertainty principle, for the reader who has sufficient back-ground for all we discussed above, i.e. GPs, FT and UR . 
       
\section{Perception, sensation und so weiter}
Hopefully equipped with the above information, we demonstrate in Fig. 5 a caricature of perception and signal conversion. Next figure (Fig. 6). shows, a simple example that ``how" a triangle is converted by FT. This is just perception received by eye cells, turned to a electrical signal and send to related \emph{center}.    

This is how we wire electromagnetic outer signal to an internal electrical one. Note that the game is not over, yet. To have a sensation from information is much more complicated. Knowledge is even more disgusting, which will be discussed in forthcoming papers. Remind yourself, we have shown in Fig. 1. which are the imposed images extracted from GS, and presented by \cite{WAGEMANS:12}. In Fig. 1, lower panels are just the Fourier transformed images. This is what we have in our brains as an electrical signal, if some has one. However, it is just the beginning, the center starts to compare this signal with previous perceptions and \emph{sensations}. Analyzed similarly, before.

At this point, to keep reader with us, we summarize: An independent signal (here EM) data reached to our cells (eye). Then converted information to an electrical signal and related cells made a FT. Next, connected brain center captured the electrical impulse and compared with previous signal store (memory). Now, the sense is there electrical, compared and filtered. Here comes \emph{knowledge}, after comparison brain produces a new information composed of previous and recent data. However, it is not finalized yet. While, with all the previous information-knowledge Etc. it is the time to react.

\section{CONCLUSION}

The long standing debate on the explanation of observer and observed in terms of \emph{seeing} is discussed with in Gestalt Principles utilizing Fourier transformation. Finally we related, GPs to uncertainty relation.  
Our approach is promising in couple of senses: It bases on firm mathematical grounds, known physiologic and art principles moreover supported by physics.  

\section*{Acknowledgements}
This work was not financially supported, other than personal fundings.
\begin{figure}
	\includegraphics[width=0.9\columnwidth]{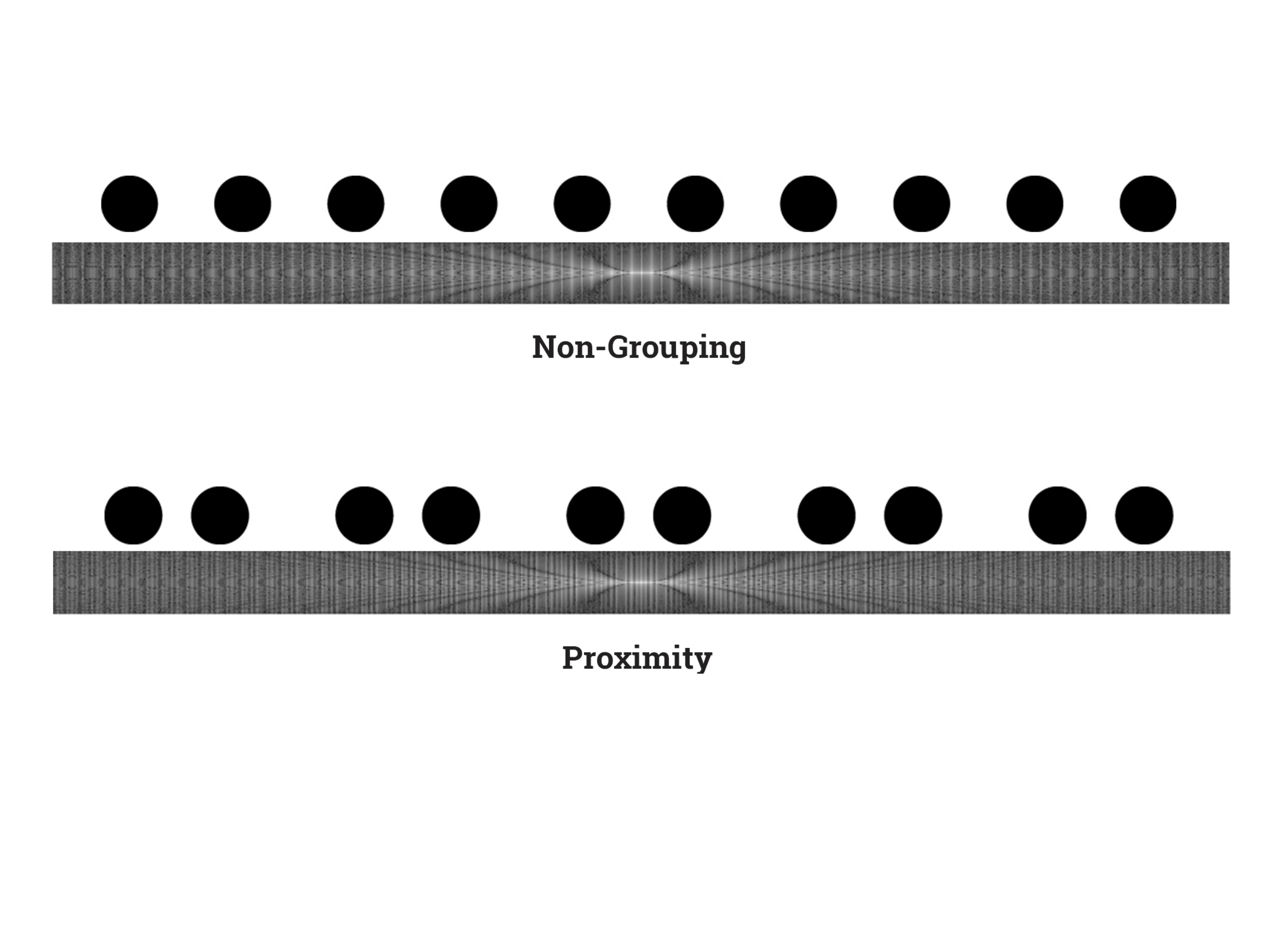}
	\caption{(color online)  \label{fig1} Equally spaced black dots on white background, upper most panel, and the FT of regarding distribution. Lower panel, coupled black dots and their FT (lower most, figure. At a first glance the FTs look very similar, however, the vertical patterns differ, in their sharpness. 
	}
\end{figure}
\begin{figure}
	\includegraphics[width=0.9\columnwidth]{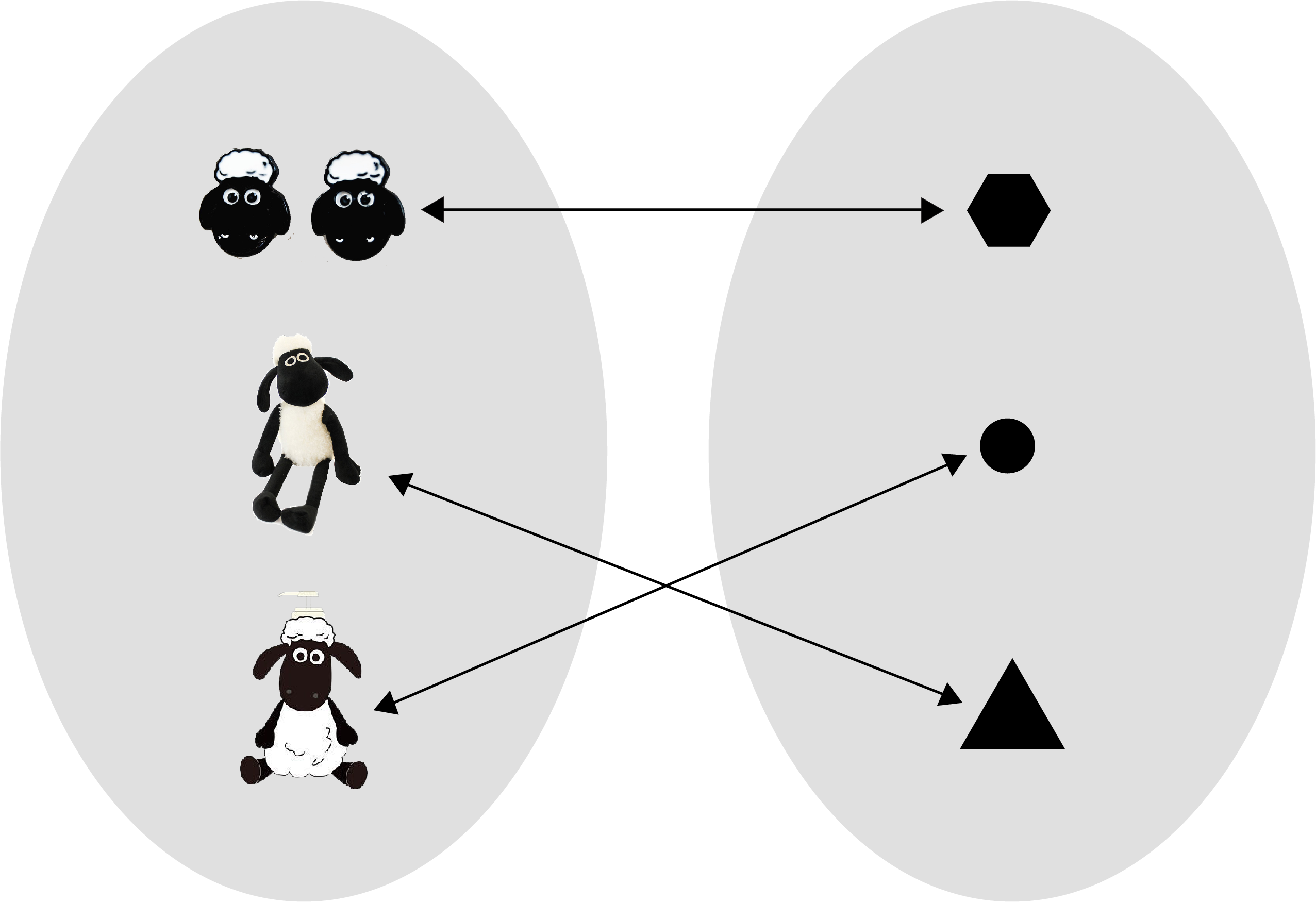}
	\caption{(color online)  \label{fig2} Two sets, two worlds. On left hand side, lambs which might be more than one (upper most element). On right hand side, some shapes which can be mapped to numbers. 
	}
\end{figure}
\begin{figure}
	\includegraphics[width=0.9\columnwidth]{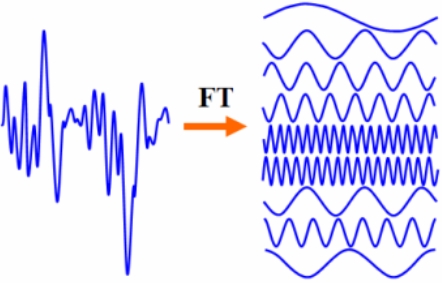}
	\caption{(color online)  \label{fig3} FT of a function (after \cite{FT:14}), which clearly demonstrates that any signal can be expressed as superposition of many known functions. Here, Sines and Cosines with period $a$.  
	}
\end{figure}
\begin{figure}
	\includegraphics[width=0.9\columnwidth]{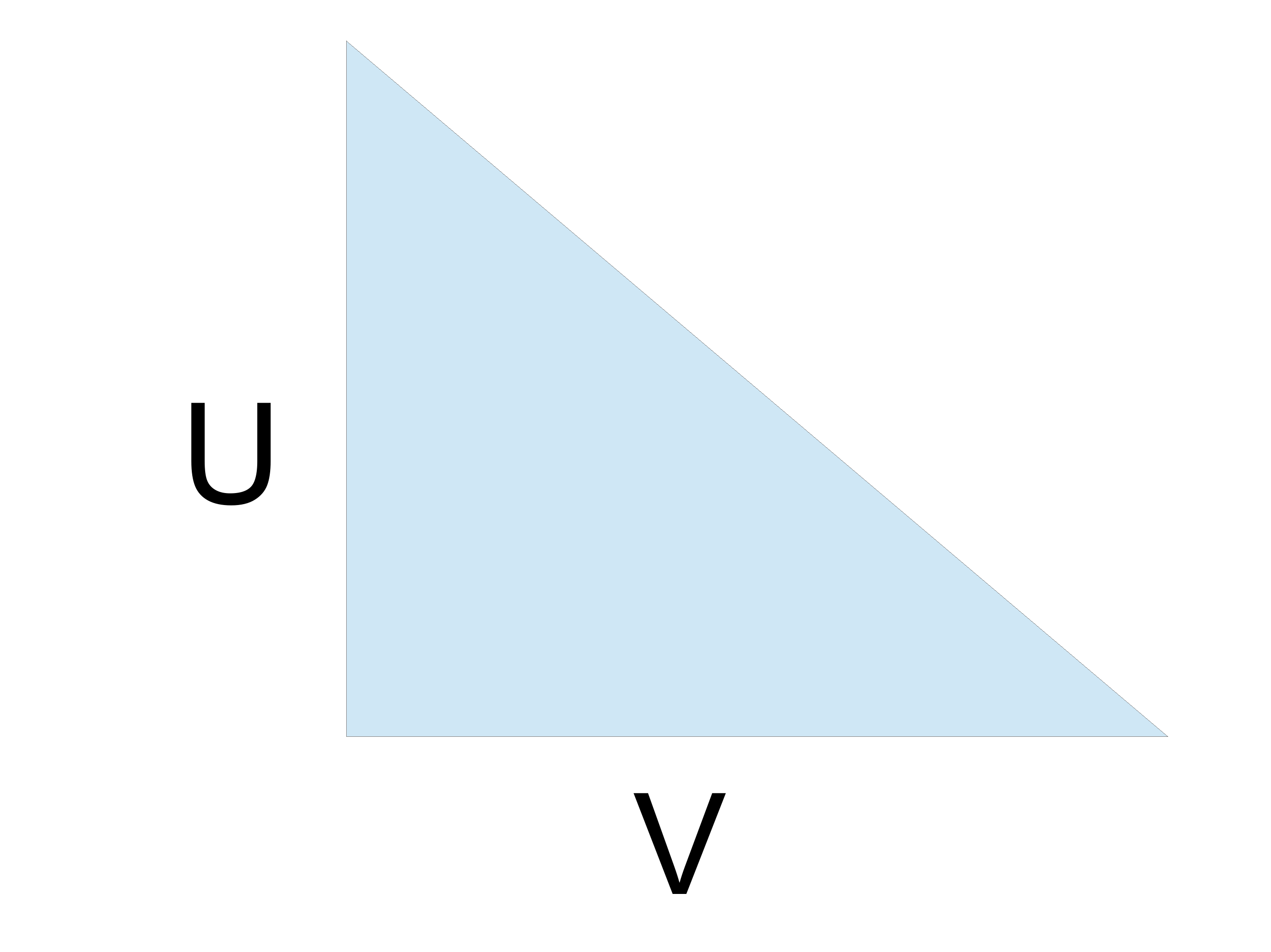}
	\caption{(color online)  \label{fig4} A well known form, right angled triangle. Due to Pythagoras theorem $v^2+u^2=$ \emph{hypotenuse}. However, it is only true and lower bound, if the dimension is two and surface is flat. Otherwise, it can be larger. As in QM. 
	}
\end{figure}
\begin{figure}
	\includegraphics[width=0.9\columnwidth]{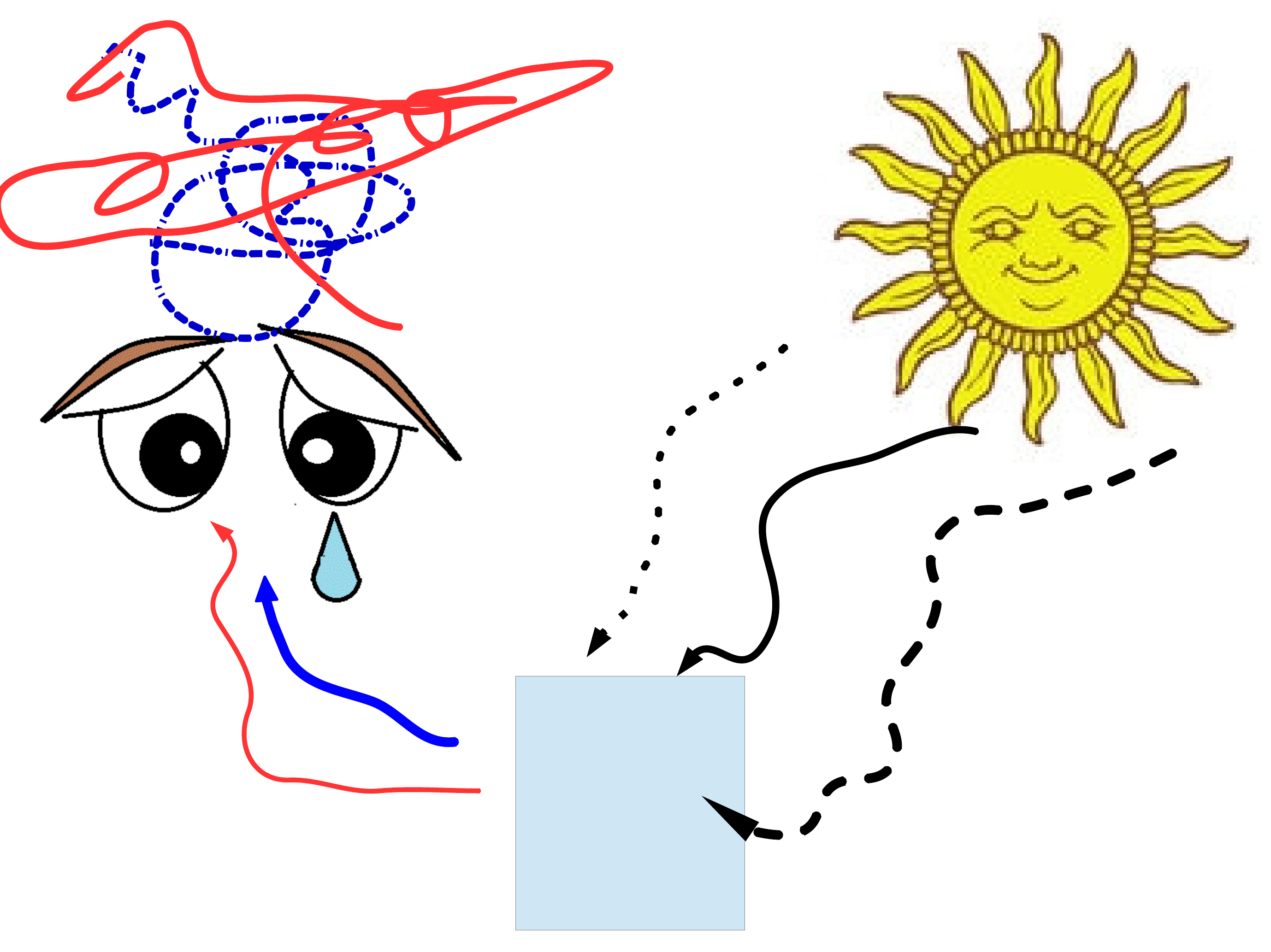}
	\caption{(color online)  \label{fig5} Caricature of Sensation. First, a light from (EM radiation in optical interval) hits the object. Next, the radiation is reflected to the observer. Note that, rest of the radiation is absorbed by the object. Cleaned up signal is received by observing cells and EM radiation is converted to electrical signals, by FT.
	}
\end{figure}
\begin{figure}
	\includegraphics[width=0.9\columnwidth]{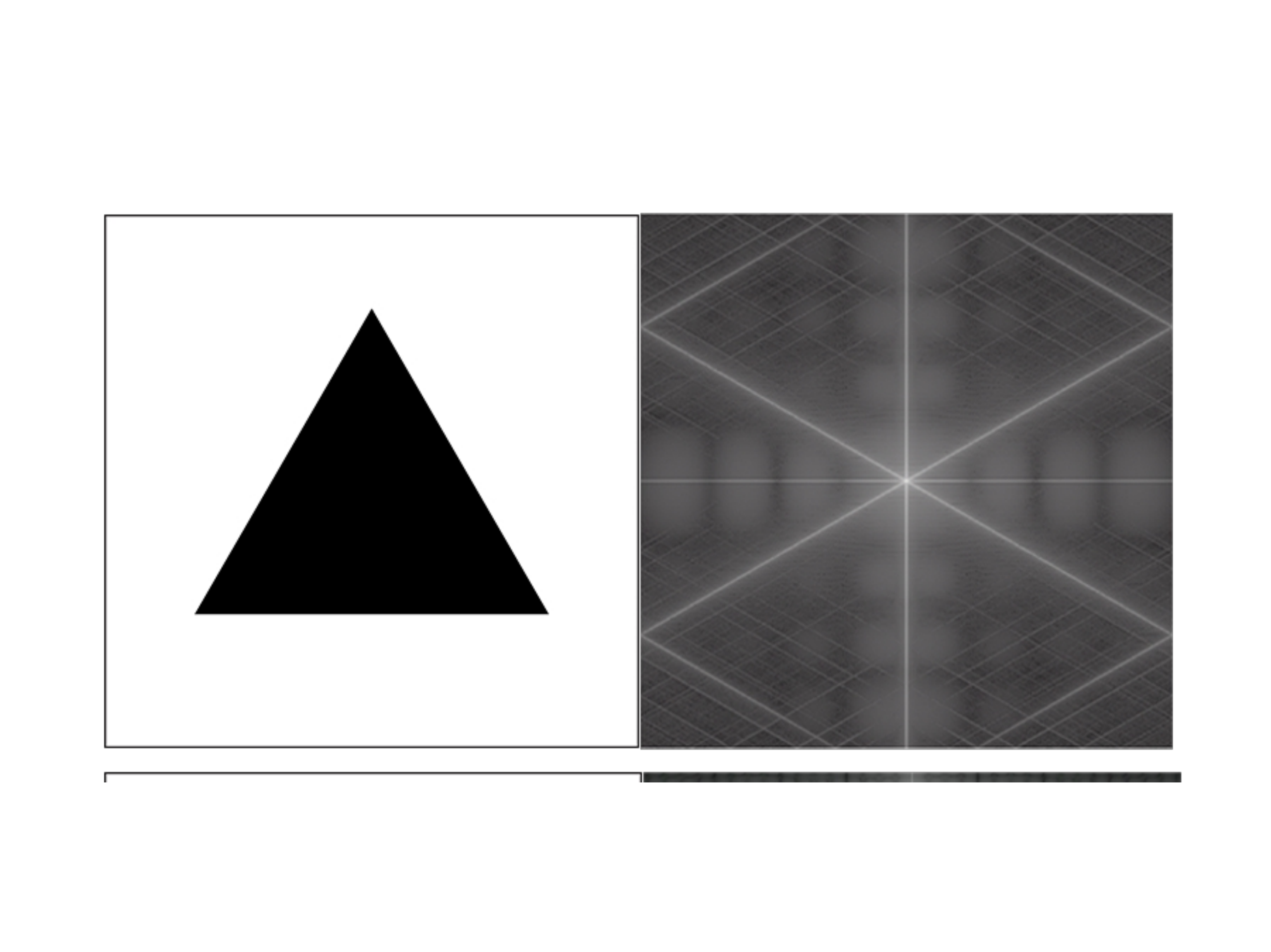}
	\caption{(color online)  \label{fig6} Left hand side, a triangular black shape or form on white background. Right hand side, its Fourier transform.   
	}
\end{figure}
\section*{References}

\bibliographystyle{elsarticle-num}

\end{document}